\documentclass{ivs2e}
\usepackage{epsfig}

\Title{How and Why to do VLBI on GPS}
\STitle{VLBI on GPS}
\PublicationType{proceedings}
\PublicationName{IVS 2010 General Meeting Proceedings}
\IvsComponentName{Hobart Network Station}
\NumberofAuthors{1}
\AuthorName{1}{J. M. Dickey}
\AuthorEmail{1}{john.dickey@utas.edu.au}
\AuthorTelephone{1}{+61-3-6226-2447}
\AuthorPersonalWebPage{1}{http://www-ra.phys.utas.edu.au/~jdickey/}
\AuthorInstitutionReference{1}{1}
\ContactAuthorReference{1}
\NumberofInstitutions{1}
\InstitutionName{1}{University of Tasmania}
\InstitutionPostAddress{1}{School of Maths and Physics\ Hobart, TAS 7005}
\InstitutionCountry{1}{Australia}
\InstitutionFax{1}{+61-3-6226-2410}
\InstitutionWebPage{1}{http://www.utas.edu.au/}
\begin{document}

\MakeTitle

\begin{abstract}
In order to establish the position of the center of mass of the Earth
in the International Celestial Reference Frame, observations of the
Global Positioning Satellite (GPS) constellation using the IVS network
are important.  With a good frame-tie between the coordinates of
the IVS telescopes and nearby GPS receivers, plus a common local
oscillator reference signal, it should be possible to observe 
and record simultaneously signals from the astrometric calibration
sources and the GPS satellites.  The standard IVS solution would
give the atmospheric delay and clock offsets to use in analysis
of the GPS data.  Correlation of the GPS signals would then give
accurate orbital parameters of the satellites {\bf in the ICRF
reference frame}, i.e. relative to the positions of the astrometric
sources.  This is particularly needed to determine motion of the
center of mass of the earth along the rotation axis.
\end{abstract}

\section{Background}

Methods of observing the Global Positioning System (GPS) satellites with the
International Very Long Baseline Interferometry Service for 
Geodesy and Astrometry (IVS) telescopes have been discussed
for more than a decade (Hase 1999 \cite{Hase_1999},
Petrachenko et al. 2004 \cite{Petrachenko_etal_2004}).  The
motivation is to measure the orbits of the GPS
spacecraft in near-real-time with high precision {\bf directly
in the reference frame} defined by the extragalactic radio sources,
i.e. the International Celestial Reference Frame
(ICRF, Ma 2008 \cite{Ma_2008}, Boboltz et al. 2010 \cite{Boboltz_2010}).
The challenge of extending the ICRF to the International
Terrestrial Reference Frame (ITRF) involves combining
data from other sources such as Satellite Laser Ranging (SLR),
the Gravity Recovery and Climate Experiment (GRACE),
the Doppler Orbitography and Radiopositioning
Integrated by Satellite (DORIS) systems, and measurements using
GPS receivers themselves (e.g. Argus et al. 2010 \cite{Argus_etal_2010},
Tregonning et al. 2009 \cite{Tregonning_etal_2009}). 

A particularly important issue is the motion of the Earth's center of
mass (CE).  A discrepency between the ITRF2000 and ITRF2005 
suggests that the mass distribution of
the Earth is changing in such a way that the CE is moving northward
at a rate of 1.8 mm yr$^{-1}$ (Altamimi et al. 2007 \cite{Altamimi_etal_2007},
Tregonning and Watson 2009 \cite{Tregonning_Watson_2009}).
If this motion were real it would raise problems with other precision
measurements such as sea level rise and glacial isostatic adjustment
(e.g. Beckley et al. 2007 \cite{Beckley_etal_2007}).
The problem is the reference frames themselves.  While SLR, GRACE,
and DORIS are ultra-sensitive to the Earth's gravitational field,
the IVS solutions for the Earth Orientation Parameters (EOP's) 
are sensitive only to the rotation of the Earth's surface.  The frame-tie
between the ICRF and
the satellite orbits (measured in the ITRF) introduces uncertainty
in the CE position at the level of millimeters.  This frame-tie
could be simplified by correlating the GPS signals along with those
from the ICRF-defining quasars as part of the IVS operation.

The CE position could be a sensitive measure of the global average
of glacial melting, given the asymmetry of the latitudinal distribution
of land and sea between the Northern and Southern Hemispheres.  Melting
of sea ice has no impact on the CE, but melting of ice caps such
as those in Greenland and Antarctica leads to a redistribution of
the mass as the melt water adjusts to
follow an equipotential.  This glacial melting
contributes to sea level rise, and so it has been studied quantitatively
(Bahr et al. 2007 \cite{Bahr_etal_2007} and references therein); the
mass of ice whose melting gives a rise in the mean
sea level of 1 mm is $m_i=3.3\cdot 10^{14}$ kg
(Dyurgerov and Meier 2005 \cite{Dyurgerov_Meier_2005} figure 5 ff).
Comparing this to the total Earth mass gives a 
prediction for the motion of the CE along
the axis of rotation of $\dot{z}$:
\[ \dot{z} \ = \ - \ R_{\oplus} \left(\frac{f \ m_i}{2\  M_{\oplus}}\right) \
\left(\frac{\dot{s}}{{ \rm mm \ yr}^{-1}}\right) \
\simeq \ -0.4 {\rm \ mm \  yr}^{-1} \]
\noindent where $R_{\oplus}$ and $M_{\oplus}$
are the Earth radius and mass, and $f$ is a dimensionless number
between -1 and +1 given by the mass-weighted mean of the
sine of the latitude where the ice melts.
Assuming that $f \sim 1$ gives the value of -0.4 mm yr$^{-1}$
above, with the minus indicating southward motion because most glacial 
melting is (for now) in the Northern Hemisphere.  More accurate measurement
of the motion of the CE would be a valuable addition to our knowledge of
the effects of global warming (Dickey et al. 2002 \cite{Dickey_etal_2002}).
This could be accomplished by determining the orbital parameters of
the GPS satellites directly from measurements coupled with routine IVS 
observations.  Combining the two would circumvent the need for a 
series of techniques, each based on a different observable, to
establish the CE together with the ITRF itself.
The main advantage of combining IVS and GPS satellite orbit measurement
is not improved precision in the measurements, although that might
be possible, but rather the direct measurement of satellite orbits
relative to the ICRF calibrators.

\begin{figure} 
\hspace{.7in}\epsfig{file=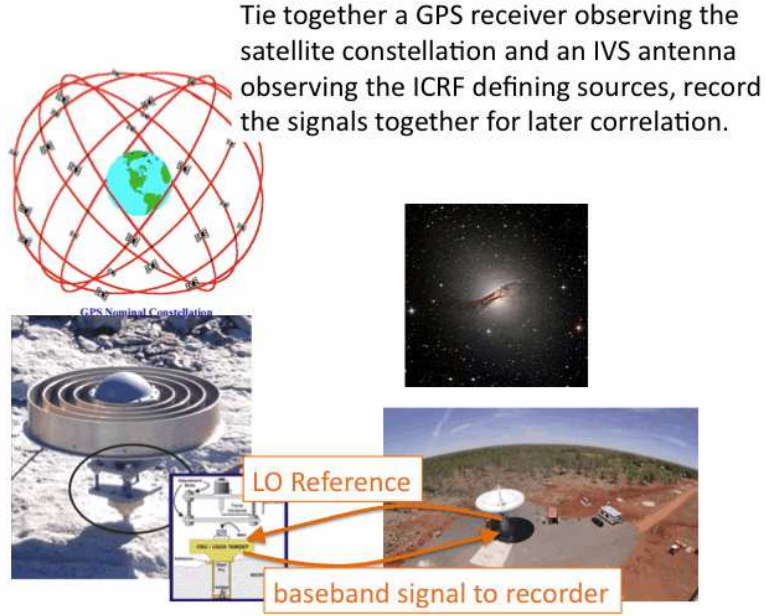,height=3.5in}

\caption{\label{fig:method_2} An illustration of the second method for
combining observations of the GPS signals with the IVS calibrators.  The
advantages of
recording signals from two antennas simultaneously at each station
outweigh the 
difficulties of measuring the offset between the two, both in position
and in the LO/RF round-trip delay.}
\end{figure}

\section{IVS Techniques}

The simplest approach (method 1) to measuring GPS signals with the IVS array is 
to use the radio telescopes themselves to record the GPS signals, and
then to correlate these signals to determine the delay on each baseline
and hence the position of each satellite as a function of time.  
The radio telescope beams are narrow, so that the satellites
are observed one at a time, and their approximate positions must be tracked
using an ephemeris.  The main disadvantage of this method is that the GPS
signals are in the 1500-1600 MHz and 1200-1300 MHz ranges, which are
well below the S-band (2100-2400 MHz) frequencies of the IVS receivers.
Thus new, wideband receivers would have to be built for all the network
telescopes, in order to accomodate observations of both GPS and the standard
IVS calibration sources together.

A more ambitious but ultimately simpler method (method 2, illustrated 
in figure \ref{fig:method_2})
is to use a standard,
geodetic-quality GPS antenna and receiver near each IVS telescope,
using local oscillators locked to the same station clock for the
GPS and IVS receivers and simultaneously recording the receiver
outputs on separate channels on the same media.  A precise frame-tie
or survey of the relative positions of the two antennas is also
needed.  The GPS antennas are nearly omni-directional and so they
receive the signals from all satellites that are above the horizon.
Correlating the outputs from the GPS receivers should give multiple delay
peaks on each baseline, one for each spacecraft.  With three or more
baselines the positions of all the satellites can be fixed, and 
their orbital parameters determined from the position vs. time
over an extended observation.  Note that it is the carrier signal that
gives the correlation peak and hence provides the relative
delay on each baseline, there is no need to decode the information carried
in the modulation of the carrier as for normal GPS operation.

The main advantage of the second method is that the radio telescopes
simultaneously carry out normal IVS observing, so that the correlation
and position solutions from the ICRF sources provide clock offsets, EOP's,
and atmospheric delays at all times during the observation.  These
results can be extended to the frequency of the GPS to predict values
for the propagation delay through the neutral and ionized components
of the atmosphere.  Determination of the clock offsets effectively
puts the GPS satellite positions in the reference frame of the astronomical
sources.  This accomplishes the frame-tie that allows the CE to be
{\bf directly measured in the ICRF} with a single technique.  
Other satellite measurement techniques could still improve the precision of
the measured CE position, but the fundamental reference frame of the
measurement would become the ICRF rather than the ITRF.  This avoids 
problems with the registration (offset) of one reference frame relative to
the other.

\section{Applications}

\begin{figure} 
\hspace{.7in}\epsfig{file=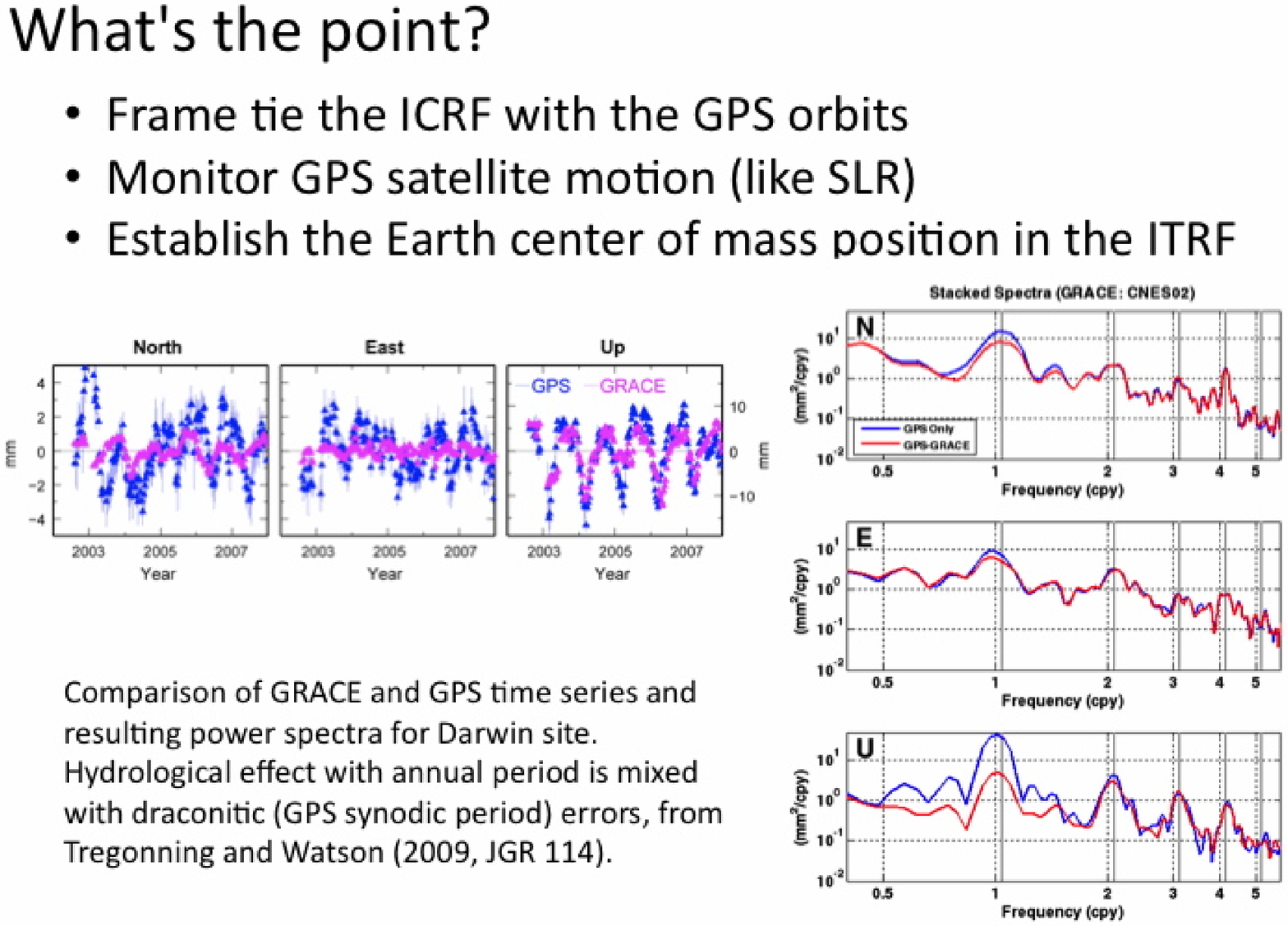,height=3.5in}

\caption{\label{fig:draconitic} 
An illustration of the current problem with GPS orbital elements.
At the draconitic period of the GPS satellites, i.e. the synodic
period of the constellation as a whole, there is a significant
discrepancy between position measurements using GRACE and those
using GPS.  The figures are taken from Tregonning and Watson (2009)
\cite{Tregonning_Watson_2009}, who studied time series of positions
for a geodetic reference site near Darwin, NT.
}
\end{figure}

Evidence of the need for better long-term monitoring of the GPS
orbital elements in the ICRF frame comes from comparison of time
series of terrestrial reference positions using GPS and GRACE.
Figure \ref{fig:draconitic} from Tregonning and Watson (2009 \cite{
Tregonning_Watson_2009}) shows the time series and resulting power
spectrum of a reference point near Darwin, NT.  There is strong
vertical motion with an annual period resulting from seasonal groundwater
variation.  Unfortunately, determination of this variation, and
hence of the groundwater supply in the region, is confused by 
another effect with a similar period, the synodic period of the
GPS constellation or draconitic period, 351.4 days.  The GRACE
and GPS positions agree well except for periods between
0.5 and 1.5 years in the vertical direction as seen on figure
\ref{fig:draconitic}.  This problem could be alleviated by 
independent measurement of the GPS satellite positions in the ICRF, 
that could be provided by the extension of the IVS technique
described here. 

\section{Conclusions}

The modest proposal of this contribution is that IVS operations
in the future should include recording of one or more bands in
the 1.5 GHz range collected from a GPS receiver near the main
radio astronomy antenna.  The GPS receiver should have its local
oscillator phase-locked to the station clock.  As a preliminary
experiment, three antennas could be equipped in this way for 
test observations.  As an ultimate goal, the GPS signal from
each satellite could be decoded to predict the delay expected
on each baseline, to save time in the correlation step.
Two very important steps toward the operation described here
are the pioneering studies by Tornatore and Haas (2010
\cite{Tornatore_Haas_2010})
and Kwak et al. (2010 \cite{Kwak_etal_2010})
described in this volume.

\section{Acknowledgements}

I am grateful to Paul Tregonning, Christopher Watson, 
Richard Coleman, and Jim Lovell for suggesting and explaining this topic
to me.

\end{document}